\documentclass[runningheads]{llncs}
\usepackage{graphicx}

\usepackage{xspace}
\usepackage{dsfont}
\usepackage{stmaryrd,amsmath}
\usepackage{semantic,multirow,multicol}
\usepackage[algoruled,linesnumbered,noend]{algorithm2e}
\usepackage[colorlinks,citecolor=blue,linkcolor=blue]{hyperref}
\usepackage[table]{xcolor}  %
\usepackage{bookmark}
\usepackage{amssymb}

\usepackage{listings}
\newcommand{\tool}{{\sc QMVerif}\xspace}

\newcommand{\Bb}{\mathbb{D}}

\newcommand{\Expr}{\mathcal{E}}

\newcommand{\Op}{\mathcal{O}}

\newcommand{\type}{\mathcal{T}}

\newcommand{\return}{{\tt return}}
\newcommand{\rud}{{\sf RUD}}
\newcommand{\sid}{{\sf SID}}

\newcommand{\ukd}{{\sf UKD}}

\newcommand{\sdd}{{\sf SDD}}
\newcommand{\Dom}{{\sf Dom}}

\newcommand{\SI}{{\bf SI}}
\newcommand{\UF}{{\bf UF}}
\newcommand{\Var}{{\sf Var}}
\newcommand{\RVar}{{\sf RVar}}

\newcommand {\sem}[1]{{\llbracket{#1}\rrbracket}}

\newcommand{\QMS}{{\tt QMS}}

\begin{document}
\title{Quantitative Verification of Masked Arithmetic Programs against Side-Channel Attacks}
\titlerunning{Verification of Masked Arithmetic Programs}


\author{Pengfei Gao\inst{1}, Hongyi Xie\inst{1}, Jun Zhang\inst{1}, Fu Song\inst{1}, Taolue Chen\inst{2}}
\institute{
School of Information Science and Technology\\
ShanghaiTech University, China
\and
Department of Computer Science and Information Systems, \\
Birkbeck, University of London, UK
}
\maketitle
\vspace{-1em}
\begin{abstract}
Power side-channel attacks, which can deduce secret data via statistical analysis, have become a serious threat. Masking is an effective countermeasure for reducing the statistical dependence between secret data and side-channel information. However, designing masking algorithms is an error-prone process. In this paper, we propose a hybrid approach combing type inference and model-counting to verify masked arithmetic programs against side-channel attacks. The type inference allows an efficient, lightweight procedure to determine most observable variables whereas model-counting accounts for completeness. In case that the program is not perfectly masked, we also provide a method to quantify the security level of the program. We implement our methods in a tool \tool and evaluate it on cryptographic benchmarks. The experimental results show the effectiveness and efficiency of our approach.
\end{abstract}

\vspace*{-1.2em}
\section{Introduction}\label{sec:intro}
Side-channel attacks aim to infer secret data (e.g. cryptographic keys) by exploiting statistical dependence between secret data and non-functional properties such as execution time~\cite{Kocher96}, power consumption~\cite{KJJ99}, and electromagnetic radiation~\cite{QS01}. They have become a serious threat in application domains such as cyber-physical systems. As a typical example, the power consumption of a device executing the instruction $c=p\oplus k$ usually depends on the secret $k$, and this can be exploited via \emph{differential power analysis} (DPA)~\cite{MBKP11} to deduce $k$. 

\emph{Masking} is one of the most widely-used and effective countermeasure to thwart side-channel attacks. Masking is essentially a randomization technique for reducing the statistical dependence between secret data and side-channel information (e.g. power consumption).
For example, using Boolean masking scheme, one can mask the secret data $k$ by applying the exclusive-or ($\oplus$) operation with a random variable $r$, yielding a masked secret data $k\oplus r$. It can be readily verified that the distribution of $k\oplus r$ is independent of the value of $k$ when $r$ is uniformly distributed. Besides Boolean masking scheme, there are other masking schemes such as additive masking schemes (e.g. $(k+r)\bmod n$) and multiplicative masking schemes (e.g. $(k\times r)\bmod n$).
%
%
A variety of masking implementations such as AES and its non-linear components (S-boxes) have been published over the years. 
However, designing effective and efficient masking schemes is still a notoriously difficult task, especially for non-linear functions.
This has motivated a large amount of work on verifying whether masked implementations, as either (hardware) circuits or (software) programs, are statistically independent of secret inputs. 
Typically, masked hardware implementations are modeled as (probabilistic) Boolean programs where all variables range over the Boolean domain (i.e. $\mathds{GF}(2)$), while masked software implementations, featuring a richer set of operations, require to be modeled as (probabilistic) arithmetic programs.

Verification techniques for masking schemes can be roughly classified into type system based approaches~\cite{MOPT12,BBDFGS15,BBDFGSZ16,BBFG18,BHL17,BMZ17,Coron18}
and model-counting based approaches~\cite{EWS14a,EWS14b,ZGSW18}.
%
%
The basic idea of type system based approaches is to infer a \emph{distribution type} for observable variables in the program that are potentially exposed to attackers. From the type information one may be able to show that the program is secure.
This class of approaches is generally very efficient mainly because of their static analysis nature. However, they may give inconclusive answers 
as most existing type systems do not provide completeness guarantees. 


Model-counting based approaches, unsurprisingly, 
encode the verification problem as a series of
model-counting problems, and typically 
leverage SAT/SMT solvers. The main advantage of this approach is its completeness guarantees. However, the size of the SMT formula is exponential in the number of (bits of) random variables used in masking, hence the approach poses great challenges to its scalability. We mention that, within this category, some work further exploits Fourier analysis~\cite{BCG13,BGIKMW18}, which considers 
the Fourier expansion 
of the Boolean functions. The verification problem can then be reduced to checking whether certain coefficients of the Fourier expansion are zero or not.
Although there is no hurdle in principle, to our best knowledge, currently model-counting based approaches are limited to Boolean programs only.

While verification of masking for Boolean programs is well-studied~\cite{EWS14a,EWS14b,ZGSW18}, generalizing them to arithmetic programs brings additional challenges. First of all, arithmetic programs admit more operations which are absent from Boolean programs. A typical example is field multiplication. In the Boolean domain, it is nothing more than the logical AND operator. However for $\mathds{GF}(2^n)$ (typically $n=8$ in cryptographic algorithm implementations), the operation is nontrivial which prohibits many optimization which would otherwise be useful for Boolean domains. Second, verification of arithmetic programs often suffers from serious  scalability issues, especially when the model-counting based approaches are applied. 
We note that transforming arithmetic programs into equivalent Boolean versions is theoretically possible, but  
suffer from several deficiencies: (1) one has to encode complicated arithmetic operations (e.g. finite field multiplication) as bitwise operations; (2) the resulting Boolean program needs to be checked against high-order attacks which are supposed to observe multiple observations simultaneously. This is a far more difficult problem. Because of this, we believe such an approach is practically unfavourable, if not infeasible.



Perfect masking is ideal but not necessarily holds when there are flaws or only a limited number of random variables are allowed for efficiency consideration. 
In case that the program is not perfectly masked (i.e., a potential side channel does exist), naturally one wants to tell how severe it is. For instance, one possible measure is 
the resource the attacker needs to invest in order to infer the secret from the side channel. For this purpose, we adapt the notion of \emph{Quantitative Masking Strength}, with which a correlation of the number of power traces to successfully infer secret data has been established empirically~\cite{EWTS14,EWTS15}.
%
%

\smallskip
\noindent
{\bf Main contributions.} We mainly focus on the verification of masked \emph{arithmetic} programs. We advocate a hybrid verification method combining type system based and model-counting based approaches, and provide additional quantitative analysis. We summarize the main contributions as follows.
\vspace*{-0.6em}\begin{itemize}
\item  We provide a hybrid approach which integrates type system based and model-counting based approaches into a framework, and support a sound and complete reasoning of masked arithmetic programs.
%
%
\item We provide quantitative analysis in case when the masking is not effective, to calculate a quantitative measure of the information leakage.
%
%

%
\item We provide various heuristics and optimized algorithms to significantly improve the scalability of previous approaches. 
\item  We implement our approaches in a software tool and provide thorough evaluations. Our experiments show orders of magnitude of improvement with respect to previous verification methods on common benchmarks.
%
\end{itemize}

One of the advantages of our approaches is 
the simplicity which renders them amenable for implementations and easily extensible to other settings. We also find, perhaps surprisingly, that for model-counting, the widely adopted approaches based on SMT solvers (e.g.~\cite{EWS14a,EWS14b,ZGSW18}) may not be the best approach, as our experiments suggest that an alternative brute-force approach is comparable for Boolean programs, and significantly outperforms for arithmetic programs.


\smallskip
\noindent
{\bf Related work.} The $d$-threshold probing model 
is the de facto standard leakage model for formal verification of masked programs against  
order-$d$ power side-channel attacks~\cite{ISW03}. This paper focuses on the case that $d=1$. 
Other models like noise leakage model~\cite{CJRR99,PR13}, bounded moment model~\cite{BDFGSS17}, and threshold probing model with transitions/glitch~\cite{CGPRRV12,BGIKMW18}
could be reduced to the threshold probing model, at the cost of introducing higher orders~\cite{BBDFGS15}.
Other work on side channels such as execution-time, faults, and cache do exist (\cite{Kocher96,ABBDE16,AGHK17,BS97,EWW16,BDFGZ14,BKMO14,GWW18} to cite a few),
but is orthogonal to our work.

Type systems have been widely used in the verification of side channel attacks with early work~\cite{MOPT12,BRNI13}, where masking compilers are provided which can transform an input program into a functionally equivalent program that is resistant to first-order DPA. However, these systems either are limited  to certain operations (i.e., $\oplus$ and table look-up), or suffer from unsoundness and incompleteness under the threshold probing model. 
%
%
To support verification of high-order masking, Barthe et al. introduced the notion of noninterference (NI,~\cite{BBDFGS15}), 
and strong $t$-noninterference (SNI,~\cite{BBDFGSZ16}), 
%
%
which were extended to 
give a unified framework for both software and hardware implementations in {\tt maskVerif}~\cite{BBFG18}.
Further work along this line includes improvements for efficiency~\cite{BMZ17,Coron18}, generalization for assembly-level code~\cite{BGIKMW18}, 
and extensions with glitches for hardware programs~\cite{FGPPS17}. As mentioned earlier, these approaches are incomplete, i.e., secure programs may fail to pass their verification.

\cite{EWS14a,EWS14b} proposed a model-counting based approach for Boolean programs by leveraging SMT solvers, which is complete but limited in scalability.
To improve efficiency, a hybrid approach integrating type-based and model-counting based approaches~\cite{EWS14a,EWS14b} was proposed in~\cite{ZGSW18},
which is similar to the current work in spirit. However, it is limited to Boolean programs and qualitative analysis only.
\cite{EWTS14,EWTS15} extended the approach of~\cite{EWS14a,EWS14b} for quantitative analysis, but is limited to Boolean programs.
The current work not only extends the applicability but also achieves significant improvement in efficiency even for Boolean programs (cf.\ Section~\ref{sec:exper}).
We also find that solving model-counting via SMT solvers~\cite{EWS14b,ZGSW18} may not be the best approach, in particular for arithmetic programs.

Furthermore, we mention that masking synthesis is recently proposed \cite{EWW16,BBDFGSZ16} to transform an input program into a functionally equivalent, perfectly masked one. This technique is based on the perfect masking verification~\cite{EWS14a,EWS14b,BBDFGS15}.
Our work is also related to quantitative information flow (QIF)~\cite{MH10,PMPd14,VEBAH16,PM14,BEHLMQ18}
which leverages notions from information theory (typically Shannon entropy and mutual information) to measure the flow of information in programs. The QIF framework has also been specialized to side-channel analysis~\cite{PBPMB17,PPM16,MKPPL18}.
The main differences are, first of all, QIF targets fully-fledged programs (including branching and loops) so program analysis techniques (e.g. symbolic execution) are needed, while we deal with more specialized (transformed) masked programs in straight-line forms; second, to measure the information leakage quantitatively, our measure is based on the notion of QMS which is correlated with the number of power traces needed to successfully infer the secret, while QIF is based on a more general sense of information theory; third, for calculating such a measure, both work rely on model-counting. In QIF, the constraints over the input are usually linear, 
but the constraints in our setting involve arithmetic operations in rings and fields. Randomized approximate schemes can be exploited in QIF~\cite{MKPPL18,BEHLMQ18} which is not suitable in our setting. Moreover, we mention that in QIF, input variables should in principle be partitioned into public and private variables, and the former of which needs to be existentially quantified. This was briefly mentioned in, e.g.,~\cite{MKPPL18} but without implementation.

\vspace*{-0.3em}
\section{Preliminaries}\label{sec:pre}
Let us fix a bounded integer domain $\Bb=\{0,\cdots,2^n-1\}$, where $n$ is a fixed positive integer.
Bit-wise operations are defined over $\Bb$, but we shall also consider arithmetic operations over $\Bb$ which include $+, -, \times$ modulo $2^n$ for which $\Bb$ is consider to be a ring and the Galois field multiplication $\odot$ where $\Bb$ is isomorphic to $\mathds{GF}(2)[x]/(p(x))$ (or simply $\mathds{GF}(2^n)$) for some irreducible polynomial $p$.
For instance, in AES one normally uses $\mathds{GF}(2^8)$ and $p(x)=x^8 + x^4 +x^3 + x^2 + 1$.

\vspace*{-0.7em}\subsection{Cryptographic Programs}\vspace*{-0.4em}
\label{sec:prelim-1}
We focus on programs written in C-like code that implement cryptographic algorithms such as AES, as opposed to arbitrary software programs.
To analyze such programs, it is common to assume that they are given in straight-line forms (i.e., branching-free) over $\Bb$~\cite{EWS14b,BBDFGS15}.
The syntax of the program under consideration is given as follows, where $c\in \Bb$.\vspace*{-0.5em}
\[\begin{array}{lrl}
\mbox{Operation:}~ &  \Op \ni  \circ&::= \oplus \mid \wedge \mid \vee \mid  \odot \mid +\mid -\mid \times \\
\mbox{Expression:}~&          e& ::=  c \mid x \mid e\circ e \mid \neg e \mid e\ll c\mid e\gg c\\
\mbox{Statememt:} ~&          {\tt stmt}& ::=  x\leftarrow e \mid {\tt stmt}; {\tt stmt}  \\
\mbox{Program:}  ~ &          P(X_p,X_k,X_r)& ::={\tt stmt}; \return \ x_1,...,x_m;
\end{array}
\]
A program $P$ consists of a sequence of assignments followed by a return statement. An assignment $x\leftarrow e$ assigns the value of the expression $e$ to the variable $x$, where $e$ is built up from a set of variables and constants using (1) bit-wise operations \emph{negation} ($\neg$), \emph{and} ($\wedge$), \emph{or} ($\vee$), \emph{exclusive-or} ($\oplus$), \emph{left shift} $\ll$ and \emph{right shift} $\gg$; (2) modulo $2^n$ arithmetic operations: \emph{addition} ($+$), \emph{subtraction} ($-$), \emph{multiplication} ($\times$); and (3) finite-field \emph{multiplication} ($\odot$) (over $\mathds{GF}(2^n)$)\footnote{Note that addition/subtraction over Galois fields is essentially bit-wise exclusive-or.}.
We denote by $\Op^\ast$ the extended set $\Op\cup\{\ll,\gg\}$ of operations. 

Given a program $P$, let $X=X_p \uplus  X_k\uplus X_i\uplus X_r$ denote the set of variables used in $P$,
where $X_p$, $X_k$ and $X_i$ respectively denote the set of public input, private input and internal variables,
and $X_r$ denotes the set of (uniformly distributed) random variables for \emph{masking}
private variables.
We assume that the program is given in the \emph{single static assignment} (SSA) form 
and each expression uses at most one operator. (One can easily transform an arbitrary straight-line program into an equivalent one satisfying these conditions.)
For each assignment $x\leftarrow e$ in $P$, the computation $\Expr(x)$ of $x$ is an expression obtained from $e$ by iteratively replacing all the occurrences of the internal variables in $e$ by their defining expressions in $P$. SSA form guarantees that $\Expr(x)$ is well-defined.

\smallskip
\noindent \emph{Semantics}.
A \emph{valuation} is a function $\sigma:X_p\cup X_k\rightarrow \Bb$ assigning to each variable $x\in X_p\cup X_k$ a value $c\in\Bb$. Let $\Theta$ denote the set of all valuations. Two valuations $\sigma_1,\sigma_2\in\Theta$ are \emph{$Y$-equivalent}, denoted
by $\sigma_1\approx_{Y}\sigma_2$, if $\sigma_1(x)=\sigma_2(x)$ for all $x\in Y$.

Given an expression $e$ in terms of $X_p\cup X_k\cup X_r$ and a valuation $\sigma\in\Theta$,
we denote by $e(\sigma)$ the expression obtained from $e$
by replacing all the occurrences of variables $x\in X_p\cup X_k$ by their values $\sigma(x)$,
and denote by $\sem{e}_{\sigma}$ the distribution of $e$ (with respect to the uniform distribution of random variables $e(\sigma)$ may contain).
Concretely, $\sem{e}_{\sigma}(v)$ is the probability
of the expression $e(\sigma)$ being evaluated to $v$ for each $v\in \Bb$.
For each variable $x\in X$ and valuation $\sigma\in\Theta$, we denote by $\sem{x}_{\sigma}$ the distribution $\sem{\Expr(x)}_{\sigma}$.
The semantics of the program $P$ is defined as a (partial) function $\sem{P}$
which takes a valuation $\sigma\in\Theta$ and
an internal variable $x\in X_i$ as inputs, returns the distribution $\sem{x}_\sigma$
of $x$.

\subsection{Threat Models and Security Notions}
We assume that the adversary has access to public input $X_p$, but not to private input $X_k$ or random variables $X_r$, of a program $P$.
However, the adversary may have access to an internal variable $x\in X_i$ via side-channel information. Under these assumptions, the goal of the adversary is to deduce the information of $X_k$.

\begin{definition}
Let $P$ be a program. For every internal variable $x\in X_i$,
\begin{itemize}
  \item $x$ is \emph{uniform} in $P$, denoted by $x$-$\UF$, if
$\sem{P}(\sigma)(x)$ is uniform for all $\sigma \in \Theta$.
  \item $x$ is \emph{statistically independent} in $P$, denoted by $x$-$\SI$, if
$\sem{P}(\sigma_1)(x)=\sem{P}(\sigma_2)(x)$ for all $(\sigma_1,\sigma_2)\in \Theta^2_{X_p}$, where
$\Theta^2_{X_p}:=\{(\sigma_1,\sigma_2)\in\Theta\times\Theta \mid \sigma_1\approx_{X_p}\sigma_2\}$. 
\end{itemize}
\end{definition}

The following property is straightforward.
Note that its converse  does not hold in general.

\begin{proposition}\label{prop:uniform2SI}
If the program $P$ is $x$-$\UF$, then $P$ is $x$-$\SI$.
\end{proposition}

\begin{definition}
For a program $P$, a variable  $x$ is \emph{perfectly masked} (a.k.a. secure under $1$-threshold probing model~\cite{ISW03}) in $P$ if it is $x$-$\SI$, otherwise
$x$ is leaky.

$P$ is \emph{perfectly masked} if all internal variables in $P$ are perfectly masked.
\end{definition}

\subsection{Quantitative Masking Strength}
When a program is not perfectly masked, it is important to quantify how secure it is. For this purpose,
we adapt the notion of \emph{Quantitative Masking Strength} (QMS) from~\cite{EWTS14,EWTS15} to quantify the strength of masking countermeasures. 

\begin{definition}
	The \emph{quantitative masking strength} $\QMS_{x}$ of a variable $x\in X$,
	is defined as: $1-\max_{(\sigma_1,\sigma_2)\in\Theta^2_{X_p},c\in \Bb}\Big(\sem{x}_{\sigma_1}(c)-\sem{x}_{\sigma_2}(c)\Big).$

Accordingly, the quantitative masking strength of the program $P$ is defined by $\QMS_P:=\min_{x\in X_i}{\QMS_{x}}$.
\end{definition}

The notion of QMS generalizes that of perfect masking, i.e., $P$ is $x$-$\SI$ iff $\QMS_{x}=1$.
The importance of QMS has been highlighted in~\cite{EWTS14,EWTS15}
where it is empirically shown that, for Boolean programs the number of power traces needed to determine the secret key is exponential in the QMS value.
This study suggests that computing \emph{accurate} QMS values for leaky variables is highly desirable.
%

\begin{figure}[t]
\begin{lstlisting}[multicols=2,firstnumber=1]
Cube$(k,r_0,r_1)${
  $x=k\oplus r_0$;
  $x_0=x\odot x$;
  $x_1=r_0\odot r_0$;
  $x_2=x_0\odot r_0$;
  $x_3=x_1\odot x$;
  $x_4=r_1\oplus x_2$;
  $x_5=x_4\oplus x_3$;
  $x_6=x_0\odot x$;
  $x_7=x_6\oplus r_1$;
  $x_8=x_1\odot r_0$;
  $x_9=x_8\oplus x_5$;
  return $(x_7,x_9)$;
}
\end{lstlisting}
\caption{A buggy version of the cubing algorithm from~\cite{RP10}}
\label{fig:Runningexample} \vspace*{-1.5em}
\end{figure}
\begin{example}
Let us consider the program in Fig.~\ref{fig:Runningexample}, which implements a buggy cubing algorithm in $\mathds{GF}(2^8)$ from~\cite{RP10}.
Given a secret key $k$, to avoid first-order side-channel attacks, $k$ is masked by a random variable $r_0$  leading to two shares $x= k\oplus r_0$ and $r_0$.
{\tt Cube}$(k,r_0,r_1)$ returns two shares $x_7$ and $x_9$ such that $x_7\oplus x_9=k^3:=k\odot k\odot k$, where $r_1$ is another random variable.

{\tt Cube} computes $k\odot k$ by $x_0=x\odot x$ and $x_1=r_0\odot r_0$ (Lines 3-4), as $k\odot k=x_0\oplus x_1$.
Then, it computes $k^3$ by a secure multiplication of two pairs of shares $(x_0,x_1)$ and $(x,r_0)$ using the random variable $r_1$ (Lines 5-12).
However, this program is vulnerable to first-order side-channel attacks.
As shown in~\cite{RP10}, we shall refresh $(x_0,x_1)$ before computing $k^2\odot k$ by inserting
$x_0=x_0\oplus r_2$ and $x_1=x_1\oplus r_2$ after Line 4, where $r_2$ is a random variable.
We use this buggy version as a running example to illustrate our techniques.

As setup, we have: $X_p=\emptyset$, $X_k=\{k\}$, $X_r=\{r_0,r_1\}$
and $X_i=\{x,x_0,\cdots,x_9\}$. The computations $\Expr(\cdot)$ of internal variables are:

\noindent{\footnotesize
$\begin{array}{ll}
  \Expr(x)=k\oplus r_0 ~~~~~ \Expr(x_0)=(k\oplus r_0)\odot (k\oplus r_0) & \Expr(x_1)=r_0\odot r_0 \\
  \Expr(x_2)=((k\oplus r_0)\odot (k\oplus r_0))\odot r_0 &  \Expr(x_3)=(r_0\odot r_0)\odot (k\oplus r_0) \\
  \Expr(x_4)=r_1\oplus (((k\oplus r_0)\odot (k\oplus r_0))\odot r_0) & \Expr(x_6)=((k\oplus r_0)\odot (k\oplus r_0))\odot (k\oplus r_0) \\
  \multicolumn{2}{l}{\Expr(x_5)=(r_1\oplus ((k\oplus r_0)\odot (k\oplus r_0))\odot r_0)\oplus ((r_0\odot r_0)\odot (k\oplus r_0)) }\\
  \Expr(x_7)=(((k\oplus r_0)\odot (k\oplus r_0))\odot (k\oplus r_0))\oplus r_1 \ \  &   \Expr(x_8)=(r_0\odot r_0)\odot r_0  \\
  \multicolumn{2}{l}{\Expr(x_9)=((r_0\odot r_0)\odot r_0)\oplus ((r_1\oplus ((k\oplus r_0)\odot (k\oplus r_0)\odot r_0))\oplus ((r_0\odot r_0)\odot (k\oplus r_0)) )}
\end{array}$}
\end{example}


\section{Three Key Techniques}
\label{sec:3proc}
In this section, we introduce three key techniques: type system, model-counting based reasoning and reduction techniques, which will be used in our algorithm.

\subsection{Type System}
\label{sec:typesystem}
We present a type system for formally inferring \emph{distribution types} of internal variables, inspired by prior
work~\cite{OMHE17,BBDFGS15,BMZ17,ZGSW18}.
We start with some basic notations. 

\begin{definition}[Dominant variables] \label{def:dom}
Given an expression $e$, a random variable $r$ is called a \emph{dominant variable} of $e$
if the following two conditions hold:
(i) $r$ occurs in $e$ exactly once,
and (ii) each operator on the path between the leaf $r$ and the root in the abstract syntax tree of $e$ satisfies that it is either from $\{\times,\odot\}$ and one of its children is a non-zero constant or
from $\{\oplus,\neg,+,-\}$.
\end{definition}

Remark that in Definition~\ref{def:dom}, for efficiency consideration, we take a purely syntactic approach meaning that we do not simplify $e$ when checking the condition (i) that $r$ occurs exactly once.
For instance, $x$ is \emph{not} a dominant variable in $((x\oplus y)\oplus x)\oplus x$, although intuitively $e$ is equivalent to $y\oplus x$. 

Given an expression $e$, let $\Var(e)$ be the set of variables occurring in $e$, and $\RVar(e):=\Var(e)\cap X_r$.
We denote by $\Dom(e)\subseteq \RVar(e)$ the set of all dominant random variables of $e$, which
can be computed in linear time in the size of $e$. It is straightforward to have

\begin{proposition}\label{prop:dom}
If $\Dom(\Expr(x))\neq\emptyset$, then $P$ is $x$-$\UF$.
\end{proposition}

\begin{definition}[Distribution Types]
Let $\type=\{\rud,\sid,\sdd,\ukd\}$ be the set of distribution
types, where for each variable $x\in X$,
\begin{itemize}
\item $\Expr(x):\rud$ meaning that the program is $x$-$\UF$;
\item $\Expr(x):\sid$ meaning that the program is $x$-$\SI$;
\item $\Expr(x):\sdd$ meaning that the program is not $x$-$\SI$;
\item $\Expr(x):\ukd$ meaning that the distribution type of $x$ is unknown.
\end{itemize}
where $\rud$ is a subtype of $\sid$ (cf. Proposition~\ref{prop:uniform2SI}).
\end{definition}

\begin{figure}[t]
\setlength{\extrarowheight}{2.5em}	
\centering  \scalebox{0.8}{
\begin{tabular}{rrr}
    $\inference{\Dom(e)\neq\emptyset} {\vdash e:\rud}[({\sc Dom})]$
 &  $\inference{\vdash e_1\star e_2:\tau} {\vdash e_2\star e_1:\tau }[({\sc Com})]$
 &  $\inference{\vdash e:\tau} {\vdash \neg e:\tau}[({\sc Ide}$_1$)]$
 \\
    $\inference{\vdash e:\sid}{\vdash e \bullet e:\sid}[({\sc Ide}$_2$)]$
 &  $\inference{}{\vdash e \diamond e:\sid}[({\sc Ide}$_3$)]$
 &  $\inference{\vdash e:\sdd}{\vdash e \bowtie e:\sdd}[({\sc Ide}$_4$)]$
 \\
    $\inference{\Var(e)\cap X_k=\emptyset} {\vdash e:\sid}[({\sc NoKey})]$
 &  $\inference{x\in X_k} {\vdash x:\sdd}[({\sc Key})]$
 &  $\inference{\vdash e_1:\rud \ \vdash e_2:\rud \\ \Dom(e_1)\setminus\RVar(e_2)\neq \emptyset} {\vdash e_1\circ e_2:\sid}[({\sc Sid}$_{1}$)]$
 \\
    $\inference{\vdash e_1:\sid \   \vdash e_2:\sid \\ \RVar(e_1)\cap\RVar(e_2)= \emptyset} {\vdash e_1\bullet e_2:\sid}[({\sc Sid}$_{2}$)]$
 &  $\inference{ \vdash e_1:\sdd \ \vdash e_2:\rud \\ \Dom(e_2)\setminus\RVar(e_1) \neq \emptyset}  {\vdash e_1\circ e_2:\sdd }[({\sc Sdd})]$
 &  $\inference{\mbox{No rule is} \\ \mbox{appliable to } e}  {\vdash e:\ukd }[({\sc Ukd})]$
\end{tabular}}
\caption{Type inference rules, where  $\star\in\Op, \ \circ\in\{\wedge,\vee,\odot,\times\}$, $\bullet\in\Op^\ast$, $\bowtie\in\{\wedge,\vee\}$ and $\diamond\in \{\oplus,-\}$.}
\label{tab:semrules} \vspace*{-1.5em}
\end{figure}

Type judgements, as usual, are defined in the form of $\vdash e:\tau,$
where $e$ is an expression in terms of $X_r\cup X_k\cup X_p$, and $\tau\in\type$ denotes
the distribution type of $e$.
A type judgement $\vdash e:\rud$ (resp. $\vdash e:\sid$ and $\vdash e:\sdd$) is valid iff $P$ is $x$-$\UF$ (resp. $x$-$\SI$ and not $x$-$\SI$) for all variables
$x$ such that $\Expr(x)=e$.
A sound proof system for deriving valid type judgements for expressions  is given in Fig.~\ref{tab:semrules}.

Rule ({\sc Dom}) states that expression $e$ containing some dominant
variable has type $\rud$ (cf. Proposition~\ref{prop:dom}).
Rule ({\sc Com}) captures the commutative law of operators $\star\in\Op$. 
Rules ({\sc Ide}$_i$) for $i=1,2,3,4$ are straightforward.

Rule ({\sc NoKey}) states that expression $e$ has
type $\sid$ if $e$ does not use any private input.
Rule ({\sc Key}) states that each private input has type $\sdd$.

Rule ({\sc Sid}$_{1}$) states that expression
$e_1\circ e_2$ for $\circ\in\{\wedge,\vee,\odot,\times\}$ has type $\sid$,
if both $e_1$ and $e_2$ have type $\rud$, and
$e_1$ has a dominant variable $r$ which is not used by $e_2$.
Indeed, $e_1\circ e_2$ can be seen as $r\circ e_2$,
then for each valuation $\eta\in\Theta$, the distributions of $r$
and $e_2(\eta)$ are independent.
Rule ({\sc Sid}$_{2}$) states that expression
$e_1\bullet e_2$ for $\bullet\in\Op^\ast$ has type $\sid$,
if both $e_1$ and $e_2$ have type $\sid$ (as well as its subtype $\rud$), and
the sets of random variables used by $e_1$ and $e_2$ are disjoint.
Likewise, for each valuation $\eta\in\Theta$, the distributions on $e_1(\eta)$
and $e_2(\eta)$ are independent.

Rule ({\sc Sdd}) states that expression $e_1\circ e_2$ for $\circ\in\{\wedge,\vee,\odot,\times\}$
has type $\sdd$, if $e_1$ has type $\sdd$, $e_2$ has type $\rud$, and
$e_2$ has a dominant variable $r$ which is not used by $e_1$.
Intuitively, $e_1\circ e_2$ can be safely seen as $e_1\circ r$.

Finally, if no rule is applicable to an expression $e$, then $e$ has unknown distribution type. Such a type is needed because our type system is---by design---incomplete. However, we expect---and demonstrate empirically---that for cryptographic programs, most internal variables have a definitive type other than $\ukd$. As we will show later, to resolve $\ukd$-typed variables, one can resort to model-counting
(cf. Section~\ref{sec:smt}).
\begin{theorem}
If $\vdash \Expr(x):\rud$ (resp. $\vdash \Expr(x):\sid$ and $\vdash \Expr(x):\sdd$) is valid, then $P$ is $x$-$\UF$ (resp. $x$-$\SI$ and not $x$-$\SI$).
%
%
\end{theorem}

\begin{example}
Consider the program in Fig.~\ref{fig:Runningexample}, we have:
{
\[\begin{array}{llll}
  \vdash \Expr(x):\rud; & \  \vdash \Expr(x_0):\sid;  & \  \vdash \Expr(x_1):\sid;  & \  \vdash \Expr(x_2):\ukd;   \\
  \vdash \Expr(x_3):\ukd; & \  \vdash \Expr(x_4):\rud; & \ \vdash \Expr(x_5):\rud;  & \ \vdash \Expr(x_6):\ukd;  \\
  \vdash \Expr(x_7):\rud; &  \   \vdash \Expr(x_8):\sid;   & \  \vdash \Expr(x_9):\rud.  \\
\end{array}\]}
\end{example}






\subsection{Model-Counting based Reasoning}\label{sec:smt}
Recall that	for $x\in X_i$, $\QMS_{x}: = 1-\max_{(\sigma_1,\sigma_2)\in\Theta^2_{X_p},c\in \Bb} (\sem{x}_{\sigma_1}(c)-\sem{x}_{\sigma_2}(c))$.	

To compute $\QMS_{x}$, one na\"{\i}ve approach is to use brute-force to enumerate all possible valuations $\sigma$ and then to compute distributions $\sem{x}_{\sigma}$  
again by enumerating the assignments of random variables.
This approach is exponential in the number of (bits of) variables in $\Expr(x)$.

Another approach is to lift the SMT-based approach~\cite{EWTS14,EWTS15} from Boolean setting to the arithmetic one.
We first consider a ``decision" version of the problem, i.e., checking whether
$\QMS_x\geq q$ for a given rational number $q\in [0, 1]$.
It is not difficult to observe that this can be reduced to checking the satisfiability of the following logic formula:
\begin{equation} \label{eq:smtqms}
\exists \sigma_1,\sigma_2\in\Theta^2_{X_p}. \exists c\in\Bb.\big(\sharp(c=\sem{x}_{\sigma_1})-\sharp(c=\sem{x}_{\sigma_2})\big) >\Delta_x^q,
\end{equation}
where $\sharp(c=\sem{x}_{\sigma_1})$ and $\sharp(c=\sem{x}_{\sigma_2})$ respectively denote
the number of satisfying assignments of $c=\sem{x}_{\sigma_1}$ and $c=\sem{x}_{\sigma_2}$, 
$\Delta_x^q=(1-q)\times 2^m$, and $m$ is the number of bits of random variables in $\Expr(x)$.

We further encode (\ref{eq:smtqms}) as a (quantifier-free) first-order formula $\Psi_x^q$ to be solved by an off-the-shelf SMT solver (e.g. Z3~\cite{MB08}):
\begin{center}
$\Psi_x^q := (\bigwedge_{f:\RVar(\Expr(x))\rightarrow\Bb}(\Theta_f\wedge \Theta_f'))\wedge \Theta_{\tt b2i}\wedge \Theta_{\tt b2i}'\wedge \Theta_{\tt diff}^q$
\end{center}
where
\begin{itemize}
  \item {\bf Program logic} ($\Theta_f$ and $\Theta_f'$): for every $f:\RVar(\Expr(x))\rightarrow\Bb$,
  $\Theta_f$ encodes $c_f=\Expr(x)$ into a logical formula with each occurrence of a random variable $r\in \RVar(\Expr(x))$ being replaced by its value $f(r)$,
  where $c_f$ is a fresh variable. There are
    $|\Bb|^{|\RVar(\Expr(x))|}$ distinct copies, but share the same $X_p$ and $X_k$. 
    $\Theta_f'$ is similar to $\Theta_f$ except that all variables $k\in X_k$ and $c_f$ are replaced by fresh variables $k'$
    and $c_f'$ respectively.
  \item {\bf Boolean to integer} ($\Theta_{\tt b2i}$ and $\Theta_{\tt b2i}'$):  $\Theta_{\tt b2i}:=\bigwedge_{f:\RVar(\Expr(x))\rightarrow\Bb} I_f= (c=c_f) \ ? \ 1: 0$.
  It asserts that for each $f:\RVar(\Expr(x))\rightarrow\Bb$, a fresh integer variable $I_f$ is $1$ if $c=c_f$, otherwise $0$.
  $\Theta_{\tt b2i}'$ is similar to $\Theta_{\tt b2i}$ except that $I_f$ and $c_f$ are replaced by $I_f'$ and $c_f'$ respectively.
  \item {\bf Different sums} ($\Theta_{\tt diff}^q)$:  $\sum_{f:\RVar(\Expr(x))\rightarrow\Bb} I_f-\sum_{f:\RVar(\Expr(x))\rightarrow\Bb}I_f' >\Delta_x^q$.
\end{itemize}

\begin{theorem}\label{thm:smt}
$\Psi_x^q$ is \emph{unsatisfiable} iff $\QMS_x\geq q$, and
the size of  $\Psi_x^q$ is polynomial in $|P|$ and exponential in
$|\RVar(\Expr(x))|$ and $|\Bb|$.
\end{theorem}

Based on Theorem~\ref{thm:smt}, we present an algorithm for computing
$\QMS_{x}$ in Section~\ref{sec:computeQMS}.
Note that the qualitative variant of $\Psi_x^q$ (i.e. $q=1$) can be used to decide whether $x$ is statistically independent by checking whether $\QMS_x=1$ holds. This will be used in Algorithm~\ref{alg:procs}. 

\begin{example} By applying the model-counting based reasoning to the program in Fig.~\ref{fig:Runningexample}, we can conclude that $x_6$ is perfectly masked, while $x_2$ and $x_3$ are leaky. This cannot be done by our type system or the ones in~\cite{BBDFGS15,BBDFGSZ16}.
	
To give a sample encoding, consider the variable $x_3$ for
$q=\frac{1}{2}$ and $\Bb=\{0,1,2,3\}$. We have that $\Psi_{x_3}^{\frac{1}{2}}$ is
\begin{center}
{\small $\begin{array}{c}
\left(\begin{array}{lclc}
 c_0=(0\odot 0)\odot (k\oplus 0)& ~\wedge~ &   c_0'=(0\odot 0)\odot (k'\oplus 0)   &~\wedge~ \\
 c_1=(1\odot 1)\odot (k\oplus 1)& ~\wedge~ &c_1'=(1\odot 1)\odot (k'\oplus 1) & ~\wedge~ \\
 c_2=(2\odot 2)\odot (k\oplus 2)& ~\wedge~ &c_2'=(2\odot 2)\odot (k'\oplus 2)& ~\wedge~ \\
 c_3=(3\odot 3)\odot (k\oplus 3)& ~\wedge~ &c_3'=(3\odot 3)\odot (k'\oplus 3)  &
\end{array}\right)\wedge \\
\left(\begin{array}{cccc}
 I_0 = (c=c_0) \ ? \ 1:0  &  ~\wedge~ &   I_1 = (c=c_1) \ ? \ 1:0  & ~\wedge~ \\
 I_2 = (c=c_2) \ ? \ 1:0 & ~\wedge~ & I_3 = (c=c_3) \ ? \ 1:0 &
\end{array}\right)\wedge \\
\left(\begin{array}{cccc}
I_0' = (c=c_0') \ ? \ 1:0  ~\wedge~  I_1' = (c=c_1') \ ? \ 1:0  & ~\wedge~ \\
I_2' = (c=c_2') \ ? \ 1:0   ~\wedge~  I_3' = (c=c_3') \ ? \ 1:0  &
\end{array}\right)\wedge \\
(I_0+I_1+I_2+I_3) - (I_0'+I_1'+I_2'+I_3') >(1-\frac{1}{2})^2
\end{array}$}
\end{center}
\end{example}

\subsection{Reduction Heuristics} \label{sec:reduction}
In this section, we provide various heuristics to reduce the size of formulae. These can be both applied to type inference and model-counting based reasoning.  

\smallskip
\noindent
{\bf Ineffective variable elimination}.
A variable $x$ is \emph{ineffective} in an expression $e$ 
if for all valuations $\sigma_1,\sigma_2\in\Theta$
such that $\sigma_1\approx_{\Var(e)\setminus\{x\}}\sigma_2$, we have $\sigma_1(e)=\sigma_2(e)$.
Otherwise, we say $x$ is \emph{effective} in $e$.
Clearly if $x$ is ineffective in $e$, then $e$ and $e[c/x]$ are equivalent for any $c\in\Bb$ while $e[c/x]$ contains less variables,
where $e[c/x]$ is obtained from
$e$ by replacing all occurrences of $x$ with $c$ .

Checking whether $x$ is effective or not in $e$ can be performed by a satisfiability checking of the logical formula:
$e[c/x]\neq e[c'/x]$. Obviously, $e[c/x]\neq e[c'/x]$ is satisfiable iff
$x$ is effective in $e$.

\smallskip
\noindent
{\bf Algebraic laws}. 
For every sub-expression $e'$ of the form $e_1\oplus e_1, e_1-e_1,e\circ 0$ or $0\circ e$ with $\circ\in\{\times,\odot,\wedge\}$ in the expression $e$,
it is safe to replace $e'$ by $0$, namely, $e$ and  $e[0/e']$ are equivalent.
Note that the constant $0$ is usually introduced by instantiating ineffective variables by $0$ when eliminating ineffective variables.

\smallskip
\noindent
{\bf Dominated Subexpression Elimination}. Given an expression $e$,
if $e'$ is an $r$-dominated sub-expression in $e$ and $r$ does not occur in $e$ elsewhere,
then it is safe to replace each occurrence of $e'$ in $e$ by the random variable $r$. 
Intuitively, $e'$ as a whole can be seen as a random variable when evaluating $e$.
Besides this elimination, we also allow to add mete-theorems specifying forms of sub-expressions $e'$ that can be replaced by a fresh variable.
For instance, $r\oplus ((2\times r)\wedge e'')$ in $e$, when the random variable $r$
does not appear elsewhere, can be replaced by the random variable $r$.

\smallskip
\noindent
{\bf Transformation Oracle}. We suppose there is an oracle $\Omega$ which, whenever possible, transforms
an expression $e$ into an equivalent expression $\Omega(e)$ such that type inference (may with above heuristics) can give a non-$\ukd$ type
to $\Omega(e)$. Such a transformation is required only in one program in our experiments. 

%
Let $\widehat{e}$ denote the expression obtained by applying the above heuristics (excluding transformation oracle) on the expression $e$.

\begin{lemma}
$\Expr(x)(\sigma)$ and $\widehat{\Expr(x)}(\sigma)$ have same distribution for any $\sigma\in\Theta$.
\end{lemma}

\begin{example}
Consider the variable $x_6$ in the program in Fig.~\ref{fig:Runningexample},
$(k\oplus r_0)$ is $r_0$-dominated sub-expression in $\Expr(x_6)=((k\oplus r_0)\odot (k\oplus r_0))\odot (k\oplus r_0)$,
then, we can simplify $\Expr(x_6)$ into $\widehat{\Expr(x_6)}=r_0\odot r_0 \odot r_0$.
Therefore, we can deduce that $\vdash \Expr(x_6):\sid$ by applying rule ({\sc NoKey}) on $\widehat{\Expr(x_6)}$.
\end{example}


\vspace*{-0.5em}\section{Overall Algorithms}\vspace*{-0.5em}

\begin{algorithm}[t]
	\SetAlgoNoLine
	\SetKwProg{myproc}{Function}{}{}
	\myproc{{\sc PMChecking}($P,X_p,X_k,X_r,X_i$)}{
		\ForEach{$x\in X_i$}{
			\If{$\vdash \Expr(x):\ukd$ is valid}{
				\If{$\vdash \widehat{\Expr(x)}:\ukd$ is valid}{
                    \If{$\Omega(\widehat{\Expr(x)})$ exists}{
                        Let $\vdash \Expr(x):\tau$ be valid for valid $\vdash \Omega(\widehat{\Expr(x)}):\tau$\;
                    }
                    \ElseIf{{\sf ModelCountingBasedSolver}$(\widehat{\Expr(x)})$={\sf SAT}}
				    {
					     Let $\vdash \Expr(x):\sdd$ be valid\;
					}
					\lElse{Let $\vdash \Expr(x):\sid$ be valid}
				}
				\lElse{Let $\vdash \Expr(x):\tau$ be valid for valid $\vdash \widehat{\Expr(x)}:\tau$}
			}
		}
	}
	\caption{Perfect masking verification}
	\label{alg:procs}
\end{algorithm}

\subsection{Perfect Masking Verification}
Given a program $P$ with the sets of public ($X_p$), secret ($X_k$), random
($X_r$) and internal $(X_i)$ variables, {\sc PMChecking}, given in Algorithm~\ref{alg:procs},
checks whether $P$ is perfectly masked or not. It iteratively traverses
all the internal variables. For each variable $x\in X_i$, it first applies
the type system to infer its distribution type.
If $\vdash \Expr(x):\tau$ for $\tau\neq\ukd$ is valid, then the result is conclusive.
Otherwise, we will simplify the expression $\Expr(x)$ and apply
the type inference to $\widehat{\Expr(x)}$.

If it fails to resolve the type of $x$ and $\mathcal{O}(\widehat{\Expr(x)})$ does not exist,
we apply the model-counting based (SMT-based or brute-force) method outlined in Section~\ref{sec:smt}, in particular, to check the expression $\widehat{\Expr(x)}$.
There are two possible outcomes: either $\widehat{\Expr(x)}$ is $\sid$ or $\sdd$.
We enforce $\Expr(x)$ to have the same distributional type as $\widehat{\Expr(x)}$ which might facilitate the inference for other expressions.


\begin{theorem}
$P$ is perfectly masked iff $\vdash \Expr(x):\sdd$ is not valid for any $x\in X_i$, when Algorithm~\ref{alg:procs} terminates.
\end{theorem}

We remark that, if the model-counting is disabled in Algorithm~\ref{alg:procs} where $\ukd$-typed variables are interpreted as potentially leaky, 
Algorithm~\ref{alg:procs} would degenerate to a sound type inference procedure that is fast and potentially more accurate than the one in~\cite{BBDFGS15}, owing to the optimization introduced in Section~\ref{sec:reduction}.

%

\vspace*{-0.5em}\subsection{QMS Computing}\label{sec:computeQMS}\vspace*{-0.5em}

After applying Algorithm~\ref{alg:procs}, each internal variable $x\in X_i$ is endowed by a distributional type of either $\sid$ (or $\rud$ which implies $\sid$) or $\sdd$.
In the former case, $x$ is perfectly masked meaning observing $x$ would gain nothing for side-channel attackers.
In the latter case, however, $x$ becomes a side-channel and it is natural to ask how many power traces are required to infer secret data from $x$ of which we have provided a measure formalized via QMS.

\begin{algorithm}[t]
	\SetAlgoNoLine
	\SetKwProg{myproc}{Function}{}{}
	\myproc{{\sc QMSComputing}{($P,X_p,X_k,X_r,X_i$)}}{
		
		{\sc PMChecking}($P,X_p,X_k,X_r,X_i$)\;
		\ForEach{$x\in X_i$}{
			\lIf{$\vdash \Expr(x):\sid$ is valid}{$\QMS_x:=1$}
			\Else{
                \lIf{$\RVar(\widehat{\Expr(x)})=\emptyset$}{
                    $\QMS_x:=0$}
                \Else{
    				${\tt low}:=0$;	${\tt high}:=2^{n\times|\RVar(\widehat{\Expr(x)})|}$\;
    				\While{${\tt low}< {\tt high}$}
    				{
    					${\tt mid}:=\lceil\frac{{\tt low}+ {\tt high}}{2}\rceil$; $q:=\frac{{\tt mid}}{2^{n\times|\RVar(\widehat{\Expr(x)})|}}$\;
    					\lIf{{\sf SMTSolver}$(\widehat{\Psi}_{x}^q)=${\sf SAT}}{ ${\tt high}:={\tt mid}-1$}
    					\lElse{${\tt low}:={\tt mid}$}
    				}
    				$\QMS_x:=\frac{{\tt low}}{2^{n\times|\RVar(\widehat{\Expr(x)})|}}$\;
                }
			}
		}
	}
	\caption{Computing QMS}
	\label{alg:ComputeQMS}
\end{algorithm}

{\sc QMSComputing}, given in Algorithm~\ref{alg:ComputeQMS},
computes $\QMS_x$ for each $x\in X_i$.
It first invokes the function {\sc PMChecking} for perfect masking verification.
For each $\sid$-typed variable $x\in X_i$, we can directly infer that
$\QMS_x$ is $1$. For each leaky variable $x\in X_i$,
we first check whether $\widehat{\Expr(x)}$ uses any random variables or not.
If it does not use any random variables, we directly deduce that $\QMS_x$ is $0$.
Otherwise, we use either the brute-force enumeration or an SMT-based binary search to compute $\QMS_x$.
The former one is trivial, hence not presented in Algorithm~\ref{alg:ComputeQMS}.
The latter one is based on the fact that
%
$\QMS_x=\frac{i}{2^{n\times|\RVar(\widehat{\Expr(x)})|}}$ for some integer $0\leq i\leq 2^{n\times|\RVar(\widehat{\Expr(x)})|}$.
Hence 
the while-loop in Algorithm~\ref{alg:ComputeQMS} executes at most
${\bf O}(n\times|\RVar(\widehat{\Expr(x)})|)$ times for each $x$.

Our SMT-based binary search for computing QMS values is different from the one proposed by Eldib et al.~\cite{EWTS14,EWTS15}.
Their algorithm considers Boolean programs \emph{only} and computes QMS values by directly binary searching the QMS value $q$ between $0$ to $1$ with a pre-defined step size $\epsilon$ ($\epsilon=0.01$ in~\cite{EWTS14,EWTS15}).
Hence, it only 
\emph{approximate} the actual QMS value and the binary search iterates ${\bf O}(\log(\frac{1}{\epsilon}))$ times for each internal variable.
Our approach works for more general arithmetic programs and 
computes the accurate QMS value.



\vspace*{-0.5em}
\section{Practical Evaluation}\label{sec:exper}
\vspace*{-0.5em}

We have implemented our methods in a tool named \tool, which uses Z3~\cite{MB08}
as the underlying SMT solver (fixed size bit-vector theory).
We conduct experiments of perfect masking verification and QMS computing on both Boolean and arithmetic programs.
Our experiments are conducted on a server with 64-bit Ubuntu 16.04.4
LTS, Intel Xeon CPU E5-2690 v4, and 256GB RAM.

\begin{table}[t]
	\setlength{\tabcolsep}{3pt}	
	\centering
	\caption{Results on masked Boolean programs for perfect masking verification.}\vspace*{-0.5em}
	\label{tab:BPPMV}
	\begin{tabular}{l|rrr|r|r|r}
		\hline
		\multirow{2}{*}{Name} & \multirow{2}{*}{$|X_i|$} & \multirow{2}{*}{$\sharp\sdd$} & \multirow{2}{*}{$\sharp$Count} &  \multicolumn{2}{c|}{\tool} &\multirow{2}{*}{{\sc SCInfer}~\cite{ZGSW18}} \\  \cline{5-6}
                              &                          &                               &                                   & SMT & B.F.  &    \\
		\hline
		P12  & 197k    & 0      &  0      &  2.9s     & \textbf{2.7s} &  3.8s       \\
		P13  & 197k    & 4.8k   & 4.8k     &  2m 8s  & \textbf{2m 6s}   &  38m 53s         \\
		P14  & 197k    & 3.2k   & 3.2k     &  1m 58s   & \textbf{1m 45s} &  42m 44s         \\
		P15  & 198k    & 1.6k   & 3.2k    &   \textbf{2m 25s}   & 2m 43s &  44m 12s         \\
		P16  & 197k    & 4.8k   & 4.8k    &   1m 50s   & \textbf{1m 38s} &   48m 20s        \\
		P17  & 205k    & 17.6k  &12.8k    &   1m 24s     & \textbf{1m 10s} &   81m 1s        \\  \hline
	\end{tabular}
\end{table}
%

\begin{table}[t]
\setlength{\tabcolsep}{2.5pt}	
\centering
	\caption{Results of masked Boolean programs for computing QMS Values.}\vspace*{-0.5em}
\label{tab:BPQMS}\scalebox{0.95}{
\begin{tabular}{c|r|rrrrr|rr|r|rrr}
\hline
\multirow{2}{*}{Name}& \multirow{2}{*}{$\sharp\sdd$}& \multicolumn{5}{c|}{SC Sniffer~\cite{EWTS14,EWTS15}}  & \multicolumn{6}{c}{\tool} \\ \cline{3-7}  \cline{8-13}
                    &           &  $\sharp$Iter    & Time  & Min & Max & Avg.  & $\sharp$Iter  & SMT & B.F. & Min & Max & Arg.  \\ \hline
P13 &4.8k& 480k  & 97m 23s & 0.00 & 1.00 & 0.98 & \textbf{ 0}    & 0 &  0& 0.00 & 1.00 & 0.98 \\
P14 &3.2k& 160k  & 40m 13s & 0.51 & 1.00 & 0.99 & 9.6k & 2m 56s & \textbf{39s} & 0.50 & 1.00 & 0.99 \\
P15 &1.6k& 80k   & 23m 26s & 0.51 & 1.00 & 1.00 & 4.8k & 1m 36s &  \textbf{1m 32s} & 0.50 & 1.00 & 1.00 \\
P16 &4.8k& 320k  & 66m 27s & 0.00 & 1.00 & 0.98 & 6.4k & 1m 40s &  \textbf{8s}& 0.00 & 1.00 & 0.98 \\
P17 &17.6k& 1440k & 337m 46s           & 0.00 & 1.00 & 0.93 & 4.8k & 51s & \textbf{1s} & 0.00 & 1.00 & 0.94 \\ \hline
\end{tabular}}
\end{table}

\subsection{Experimental Results on Boolean Programs}
We use the benchmarks from the publicly available cryptographic software implementations~\cite{EWS14a}, which consist of 17 Boolean programs (P1-P17). We conducted experiments on
P12-P17, which are the regenerations of MAC-Keccak reference code submitted to the SHA-3 competition held by NIST. (Relatively small examples P1-P11 are skipped which can be verified in less than 1 second.) P12-P17 are transformed into programs in the  straight-line form.

\smallskip
\noindent{\bf Perfect masking verification}.
Table~\ref{tab:BPPMV} shows the results of the perfect masking verification on P12-P17, where
Columns 2-4 show basic statistics, in particular, they give the number of internal variables,
leaky internal variables,
and internal variables which require model-counting based reasoning, respectively. 
Columns 5-6 respectively show the total time of our tool \tool using SMT-based and brute-force methods.
Column 7 shows the total time of the state-of-the-art tool {\sc SCInfer}~\cite{ZGSW18}.

We observe that: (1) our reduction heuristics significantly improve the performance compared with {\sc SCInfer}~\cite{ZGSW18} (generally 16--69 times faster for imperfectly masked programs; note that {\sc SCInfer} is based on SMT model-counting), and (2) the performance of the SMT-based and brute-force methods in our \tool  for verifying perfect masking of Boolean programs is largely leveled.


\smallskip
\noindent{\bf Computing QMS}. For comparison purposes, we implemented the algorithm of~\cite{EWS14a,EWS14b} for computing QMS values of leaky internal variables.
Table~\ref{tab:BPQMS} shows the results of computing QMS values on P13-P17 (P12 is excluded because it does not contain any leaky internal variable),
where Column 2 shows the number of leaky internal variables,
Columns 3-7 show the total number of iterations in the binary search (cf.\ Section~\ref{sec:computeQMS}),
time, the minimal, maximal and average of QMS values using the algorithm from~\cite{EWS14a,EWS14b}.
Similarly, Columns 8-13 show statistics of our tool \tool, in particular, Column 9 (resp. Column 10) shows the time of using SMT-based
(resp. brute-force) methods. Note that all the time reported in Table~\ref{tab:BPQMS} \emph{excludes} the time used for perfect masking checking.

We observe that (1) the brute-force method outperforms the SMT-based one significantly, and
(2) our tool \tool using the SMT-based method takes significant less iterations and time,
as our binary search step depends on the number of bits of random variables, but not a pre-defined value (e.g. $0.01$) as used in~\cite{EWS14a,EWS14b}.
In particular, the QMS values of leaky variables whose expressions contain no random variables (e.g. P13 and P17), do not need the binary search.


\begin{table}[t]
	\setlength{\tabcolsep}{2pt}	
	\centering
	\caption{Results of masked arithmetic programs, where P.M.V. denotes perfect masking verification, B.F. denotes brute-force, 12 S.F. denotes that Z3 emits segmentation fault after verifying 12 internal variables.}\vspace*{-0.5em}
	\label{tab:firstorder}
%
	\scalebox{0.9}{
		\begin{tabular}{l|ccc|c|c|c|c|c}
			\hline
			\multirow{2}{*}{Description}& \multirow{2}{*}{$|X_i|$} & \multirow{2}{*}{$\sharp\sdd$} & \multirow{2}{*}{$\sharp$Count} &  \multicolumn{2}{c|}{P.M.V.} &  \multicolumn{3}{c}{QMS}\\
			\cline{5-9}
			&                              &                         &                                  &    SMT & B.F.     &    SMT & B.F. & Value \\ \hline
			SecMult~\cite{RP10} &  11     & 0      &  0     & $\approx$0s           &   $\approx$0s   &  -  &   -    & 1                   \\
			Sbox (4)~\cite{CPRR13}        &   66    & 0      &  0     & $\approx$0s           &  $\approx$0s   & -   &    -    &  1                                   \\
			B2A~\cite{Goubin01}            &  8     & {\bf 0}      &  {\bf 1}    & 17s          & \textbf{2s}    & -   & - &1                                        \\
			A2B~\cite{Goubin01}            &  46     & 0       & 0      &  $\approx$0s         & $\approx$0s    & -   & -  &1                                      \\
			B2A~\cite{CGV14}            &   82    & 0      &   0    & $\approx$0s          &   $\approx$0s  & -   & - &   1                                      \\
			A2B~\cite{CGV14}            &  41     & 0       &    0   & $\approx$0s             &  $\approx$0s   & -   & - & 1                                      \\
			B2A~\cite{Coron17}             & 11      & {\bf 0}      &    {\bf 1}   & \textbf{1m 35s}          & 10m 59s    & -   & -  &1                                        \\
			B2A~\cite{BCZ18}            &  16     & 0      &       0& $\approx$0s          & $\approx$0s    &  -  & -  & 1                                      \\
			Sbox~\cite{RP10}           & 45      & 0      &    0   & $\approx$0s           & $\approx$0s    &  -  &  -   & 1                                     \\ \hline
			Sbox~\cite{SP06}           & 772      & {\bf 2}      &   {\bf 1}    & $\approx$0s           &  $\approx$0s   &  0.9s  &   $\approx$0s  &  0                                     \\
			$k^{3}$           & 11      & 2      &  2    &   96m 59s         & \textbf{0.2s}  & $>$4d   & 32s &     0.988                                    \\
			$k^{12}$           & 15      & 2      & 2      & 101m 34s           & \textbf{0.3s}  &  $>$4d  & 27s &  0.988                                            \\
			$k^{15}$           &  21     &  4     &  4     & 93m 27s (12 S.F.)   &  \textbf{28m 17s}  &  $>$4d  &  $\approx$64h  & 0.988, 0.980                                         \\
			$k^{240}$           & 23      &4       & 4     & 93m 27s (12 S.F.)   &  \textbf{30m 9s} &  $>$4d  &   $\approx$64h  & 0.988, 0.980                                        \\
			$k^{252}$           & 31      &4      & 4      & 93m 27s  (12 S.F.)   & \textbf{32m 58s}  &  $>$4d  &  $\approx$64h  &  0.988, 0.980                                    \\
			$k^{254}$           & 39      & 4      &4      & 93m 27s (12 S.F.)    & \textbf{30m 9s}   &  $>$4d  &  $\approx$64h   &  0.988, 0.980                                       \\
			\hline
	\end{tabular}}
\end{table} 

\vspace*{-0.5em}\subsection{Experimental Results on Arithmetic Programs}\vspace*{-0.5em}
We collect arithmetic programs which represent non-linear functions of masked cryptographic software implementations from literature. In Table~\ref{tab:firstorder}, Column 1 lists the name of the functions under consideration, where $k^{3},\ldots, k^{254}$ are buggy fragments of first-order secure exponentiation~\cite{RP10} without the first RefreshMask function. Columns 2-4 show basic statistics. For all experiments, we set $n=8$ and thus $\Bb=\{0,\cdots,2^8-1\}$.

\smallskip
\noindent{\bf Perfect masking verification}. Columns 5-6 in Table~\ref{tab:firstorder}  show the results of the  perfect masking verification on the programs using SMT-based and brute-force methods respectively.

We observe that: (1) some $\ukd$-typed variables (e.g. in B2A~\cite{Goubin01}, B2A~\cite{Coron17} and Sbox~\cite{SP06}, meaning that the type inference is inconclusive in these cases) can be resolved by model-counting (resulting in $\sid$-type),
and (2) on the programs (except  B2A~\cite{Coron17}) where the model-counting based reasoning is required (i.e., $\sharp$Count is non-zero), the brute-force method is significantly faster than the SMT-based one. In particular, for programs $k^{15},\ldots, k^{254}$, Z3 crashed with segmentation fault after verifying 12 internal variables in 93m, while the brute-force method comfortably returns the result. To further explain the performance of these two classes of methods, we manually examine these programs and find out that the expressions of the $\ukd$-typed variable 
in B2A~\cite{RP10} (where the SMT-based method is faster) only use exclusive-or ($\oplus$) operations and one subtraction ($-$) operation, while the expressions of the other $\ukd$-typed variables (where the brute-force method is faster) involve finite field multiplication ($\odot$).

We remark that transformation oracle and meta-theorems are only used for A2B~\cite{Goubin01}.
Theoretically, model-counting based reasoning could verify A2B~\cite{Goubin01}.
However, in our experiments both SMT-based and brute-force methods failed to terminate in 3 days,
though the brute-force method had verified more internal variables. For instance, on the expression $((2\times r_1) \oplus (x - r) \oplus r_1) \wedge r$ where $x$ is a private input and $r,r_1$ are random variables,
Z3 could not terminate in 2 days, while the brute-force method successfully verified
in a few minutes. We also tested the SMT solver Boolector~\cite{NPB15} (the winner of SMT-COMP 2018 on QF-BV, Main Track),
which failed to terminate in 3 days. Undoubtedly more systematic experiments are required in the future, but
our results suggest that, contrary to the common belief, currently SMT-based approaches are not promising, which calls for more scalable techniques. 

\smallskip
\noindent{\bf Computing QMS}.
Columns 7-9 in Table~\ref{tab:firstorder} show the results of computing QMS values of leaky variables, where Column 7 (resp. Column 8) shows the time of the SMT-based (resp. brute-force) method for computing QMS values (\emph{excluding} the time for perfect masking checking) and Column 9 shows the QMS values of all leaky variables (note that duplicated values are omitted).

We observe that: (1) the brute-force method can quickly compute the QMS values of the leaky variables in Sbox~\cite{RP10}, $k^{3}$ and $k^{12}$, but takes roughly 64 hours on the other programs, (2) surprisingly, the SMT-based method is only able to compute the QMS value of the leaky variable in Sbox~\cite{RP10}, but fails 
for the others after 4 days. Indeed, Z3 cannot even finish the first iteration of the binary search
on the smallest formula in 4 days. This, again, indicates the ineffectiveness of current SMT-based approaches.
%
We manually examine $k^3,...,k^{254}$ programs and find out that
(1) variables used in the computations $\Expr(x)$ of leaky variables $x$ are the same,
and (2) the computations that can be quickly verified contain at most 4 operations,
while the others contain at least 19 operations.




\section{Conclusion}
\label{sec:concl}

We have proposed a hybrid approach combing type inference and model-counting to verify masked arithmetic programs against first-order side-channel attacks. The type inference allows an efficient, lightweight procedure to determine most observable variables whereas model-counting accounts for completeness, bringing the best of two worlds.
We also provided model-counting based methods to quantify the amount of information leakage via side channels. 
We have presented the tool support \tool which has been evaluated on standard cryptographic benchmarks. The experimental results showed that our method significantly outperformed state-of-the-art techniques in terms of both accuracy and scalability.

Future work includes further improving SMT based model-counting techniques which currently provide no better, if not worse, performance than the na\"{i}ve brute-force method. Furthermore, generalizing the work in the current paper to verification of higher-order masking schemes remains to be a very challenging task.


\newpage

\bibliographystyle{abbrv}

\end{document}